\documentclass[english,prb,preprint]{revtex4}
\usepackage[T1]{fontenc}
\usepackage[latin9]{inputenc}
\makeatletter
\@ifundefined{textcolor}{}
{%
 \definecolor{BLACK}{gray}{0}
 \definecolor{WHITE}{gray}{1}
 \definecolor{RED}{rgb}{1,0,0}
 \definecolor{GREEN}{rgb}{0,1,0}
 \definecolor{BLUE}{rgb}{0,0,1}
 \definecolor{CYAN}{cmyk}{1,0,0,0}
 \definecolor{MAGENTA}{cmyk}{0,1,0,0}
 \definecolor{YELLOW}{cmyk}{0,0,1,0}
}

\makeatother

\usepackage{babel}
\begin{document}

\title{The Limits of Entanglement}

\author{Chad Orzel}

\affiliation{Union College Deartment of Physics and Astronomy, Schenectady, NY
12308}
\begin{abstract}
Quantum entanglement is one of the most intriguing phenomena in physics,
but many presentations of the subject leave a false impression that
it provides a sort of ``remote control'' for changing the state
of a distant particle by local manipulation of its entangled partner.
We discuss a simple example, suitable for undergraduate quantum mechanics
classes, showing that this is false, and demonstrating the limits
of entanglement.
\end{abstract}
\maketitle
The idea of quantum entanglement was introduced very early in the
development of quantum mechanics, most clearly in the famous Einstein,
Podolsky, and Rosen paper of 1935\cite{EPR}, but closely related
ideas were involved in the famous debates between Bohr and Einstein
at the 1927 and 1930 Solvay conferences. The full implications of
the idea weren't worked out until John Bell showed in 1964 that EPR-type
entanglement implied correlations between the states of widely separated
particles that are fundamentally non-local\cite{Bell}. Non-local
correlations in EPR-type systems was first demonstrated experimentally
by Freedman and Clauser in 1972\cite{ClauserEPR}, then in a series
of experiments by Alain Aspect in 1981-2\cite{Aspect,AspectTimeVar},
and numerous other experimental and theoretical treatments have continued
to refine our understanding of EPR and entanglement\cite{ReidEPRReview}.
In recent years, development of parametric downconversion sources
has even brought these experiments within reach of a well-equipped
undergraduate laboratory\cite{GalvezLabs}.

The notion of non-local correlations of the type described by EPR
and Bell is sufficiently unintuitive to seem almost magic, and as
a result, entanglement has captured the popular imagination like few
other aspects of quantum theory. The issues involved are sufficiently
subtle, though, that many attempts at understanding the phenomenon
cross the line between science and the supernatural\cite{Kaiser}.
Numerous attempts have been made to devise a system for superluminal
communication using entangled particles\cite{Herbert}, and even to
use entanglement as an explanation for psychic phenomena\cite{kooks}
or alternative medicine\cite{Milgrom}. As Kaiser argues\cite{Kaiser},
debunking these arguments, particularly Ref.\cite{Herbert}, helped
spur research into quantum foundations leading to developments like
no-cloning\cite{no-cloning} and no-signaling\cite{no-signaling,BrussNoSignalApprox}
theorems, direct experimental tests\cite{DeAngelisSuperluminalExpt},
and a deeper understanding of quantum information and relativity\cite{PeresRelatvityInformation}.
Still, misconceptions regarding the nature of entanglement persist,
and remain a source of frustration for many physicists.

One particularly tenacious misconception stems from mistaking the
correlation between the outcomes of measurements on an entangled pair
of particles for an absolute connection between the states of two
particles. Entanglement, in this view, provides a sort of ``remote
control,'' by which manipulations of the state of one particle are
instantaneously reflected in changes of the state of its entangled
partner an arbitrary distance away. This derives from statements of
the form\textbackslash{} ``the measurement of one particle instantaneously
determines the state of the other,'' which are common in introductory
discussions of EPR and entanglement. Popular treatments sometimes
take this to absurd extremes, as in a 2012 article whose author imagines
applying ``quantum'' physics to golf, and being able to direct the
path of a ball in flight by manipulating an entangled partner ball
back on the tee\cite{BovaGolf}. This misconception also underlies
most invocations of entanglement as an explanation for psychic phenomena,
through the claim that all particles were once in the same position,
and thus must be entangled, allowing the manipulation of particles
in a psychic's brain to alter the states of other objects\cite{kooks}.
Similar arguments have been used to explain ``alternative medicine''
techniques such as homeopathy\cite{Milgrom}, an area which is genuinely
problematic as it promotes the use of medically dubious remedies.
This even trips up some authors who ought to know better, as confusion
about entanglement was central to the Internet controversy over statements
made by Brian Cox\cite{ButterworthCox} in promoting his book with
Jeff Forshaw\cite{CoxForshaw}.

The notion of entanglement as a remote control for a distant particle
can easily be shown to be false by careful consideration of a simple
example, which could easily be used when the idea is first introduced
in an undergraduate course, or directly demonstrated using the apparatus
of Ref.\cite{GalvezLabs}. To be concrete, we will consider the case
of two polarization-entangled photons, though a similar argument will
work for other sorts of entangled systems. 

We begin with two photons, A and B, entangled so that they have opposite
polarizations, sent to widely separated polarization-sensitive detectors.
Using vertical ($|V\rangle$) and horizontal ($|H\rangle$) polarizations
as the basis states, we can write:

\begin{equation}
\Psi_{HV}=\frac{1}{\sqrt{2}}(|V\rangle_{A}|H\rangle_{B}-|H\rangle_{A}|V\rangle_{B})\label{eq:initial_linear}
\end{equation}
This is the classic example of a maximally entangled state, as a measurement
of the state of photon A allows one to predict with certainty the
state of photon B, no matter where it is located. If we detect photon
A with vertical polarization, photon B will always be horizontally
polarized, and vice versa.

We can, of course, transform this state into another basis, for example
using left- or right-hand circular polarization:

\begin{equation}
|R\rangle=\frac{1}{\sqrt{2}}(|V\rangle+i|H\rangle)\label{eq:RHC}
\end{equation}

\begin{equation}
|L\rangle=\frac{1}{\sqrt{2}}(|V\rangle-i|H\rangle)\label{eq:LHC}
\end{equation}
Re-writing the initial state $\Psi$ in the new basis, we find that
the entanglement is exactly preserved:

\begin{equation}
\Psi_{RL}=\frac{1}{\sqrt{2}}(|R\rangle_{A}|L\rangle_{B}-|L\rangle_{A}|R\rangle_{B})\label{eq:initial_circ}
\end{equation}
When we detect Photon A with right-hand circular polarization, photon
B will always have left-hand circular polarization, and so on.

To explore the idea of entanglement as remote control, we consider
a simple modification: inserting a quarter-wave plate before the detector
for photon A. The waveplate rotates the state vectors from one basis
into the other, so $|V\rangle\rightarrow|R\rangle$ and $|H\rangle\rightarrow|L\rangle$.
The idea of entanglement as a remote control would hold that rotating
the state of photon A should produce a corresponding rotation in photon
B. That is, by rotating the state of photon A from $|V\rangle$ to
$|R\rangle$, the state of photon B should rotate from $|H\rangle$
to $|L\rangle$, preserving the correlation between states.

After the waveplate insertion, the state of the two-photon system
is:

\begin{equation}
\Psi_{rot}=\frac{1}{\sqrt{2}}(|R\rangle_{A}|H\rangle_{B}-|L\rangle_{A}|V\rangle_{B})\label{eq:rot_psi}
\end{equation}
re-writing this in the circular polarization basis, we have:

\begin{eqnarray}
\Psi_{rot} & = & \frac{1}{\sqrt{2}}\left[|R\rangle_{A}\frac{-i}{\sqrt{2}}\left(|R\rangle_{B}-|L\rangle_{B}\right)-|L\rangle_{A}\frac{1}{\sqrt{2}}\left(|R\rangle_{B}+|L\rangle_{B}\right)\right]\nonumber \\
 & = & \frac{1}{2}\left(-i|R\rangle_{A}|R\rangle_{B}+i|R\rangle_{A}|L\rangle_{B}-|L\rangle_{A}|R\rangle_{B}-|L\rangle_{A}|L\rangle_{B}\right)\label{eq:rot_circ}
\end{eqnarray}
This state includes all four possible combinations of polarizations
for A and B, and thus will not produce the correlations characteristic
of an entangled state. When we detect photon A with right-hand circular
polarization, photon B is equally likely to have either right-hand
or left-hand circular polarization.

We can also look at the effect of the polarizer on measurements in
original the $|V\rangle-|H\rangle$ basis, where we find

\begin{eqnarray}
\Psi_{rot} & = & \frac{1}{\sqrt{2}}\left[\frac{1}{\sqrt{2}}\left(|V\rangle_{A}+i|H\rangle_{A}\right)|H\rangle_{B}-\frac{1}{\sqrt{2}}\left(|V\rangle_{A}-i|H\rangle_{A}\right)|V\rangle_{B}\right]\nonumber \\
 & = & \frac{1}{2}\left(|V\rangle_{A}|H\rangle_{B}-i|H\rangle_{A}|H\rangle_{B}-|V\rangle_{A}|B\rangle_{B}+i|H\rangle_{A}|V\rangle_{B}\right)\label{eq:rot_lin}
\end{eqnarray}
Again, after the state rotation, the correlations characteristic of
entanglement are destroyed. When we detect photon A with vertical
polarization, photon B is equally likely to have either horizontal
or vertical polarization. It is clear, then, that manipulation of
the state of photon A has not produced a corresponding change in the
state of photon B.

Looking at the results of Eq.\ref{eq:rot_circ} and Eq.\ref{eq:rot_lin},
one might be tempted to say that inserting the quarter-wave plate
has destroyed the initial entanglement, but this would be an overstatement
(albeit in the opposite direction from the original exaggerated claims).
Inspection of Eq.\ref{eq:rot_psi} shows that the correlation between
the states of photons A and B remains, provided the measurements of
the two polarizations are made in different bases. When photon A is
found to have right-hand circular polarization, photon B will always
be found to have horizontal polarization, and vice versa.

While this example uses polarization states for simplicity, similar
arguments will hold for any pair of entangled particles: electron
spins, qubit states of atoms or ions, or even continuous variables
such as position or momentum. A local modification of the state of
one particle changes the measurement bases needed to observe non-local
correlations, but does not directly modify the state of the entangled
partner. 

Entanglement between states, once established, is very robust, provided
one chooses the appropriate measurement bases. In a narrow technical
sense, then, there is some truth to the seemingly absurd claim that
two arbitrarily chosen particles may be entangled by virtue of having
been close together shortly after the Big Bang. Observing such entanglement,
however, let alone exploiting it in some paranormal manner, would
require complete knowledge of all state-rotating interactions for
each of the two particles over the intervening 13.7 billion years,
so as to choose the correct measurement bases to reveal the correlation.

The derivation of equations \ref{eq:rot_circ} and \ref{eq:rot_lin}
is well suited to class discussion or a homework assignment in an
undergraduate quantum mechanics course. Discussion of this scenario
can both help head off some common misconceptions about entanglement,
and also illuminate some of the subtle issues that make entanglement
and non-locality such a fascinating topic of study. For the philosophically
inclined, this can also provide an entry point for discussions of
different versions of quantum mechanics\cite{StyerFormulations};
while the final results will be the same for all, the underlying process
will be decribed in different very terms depending on whether the
wavefunction is viewed as a real object or merely a description of
our knowledge about the state of the system.
\begin{acknowledgments}
A version of this argument was originally presented on my blog (http://scienceblogs.com/principles/2012/03/14/entanglement-is-not-that-magic/).
Thanks to Matt Leifer for helpful discussions.\end{acknowledgments}

\end{document}